\documentclass[lettersize,journal]{IEEEtran}
\usepackage{amsmath,amsfonts}
\usepackage{algorithmic}
\usepackage{algorithm}
\usepackage{array}

\usepackage[caption=false,font=footnotesize,labelfont=sf,textfont=sf]{subfig}
\usepackage{textcomp}
\usepackage{stfloats}
\usepackage{url}
\usepackage{verbatim}
\usepackage{graphicx}
\usepackage{cite}

\begin{document}

\title{Light Field Super-Resolution: A Critical Review on Challenges and Opportunities}

\author{Sumit Sharma 
\thanks{$^1$Department of Electrical Engineering,
Indian Institute of Technology Madras, Tamil Nadu 600036, India
(e-mail: ee20d042@smail.iitm.ac.in}

}
\maketitle

\begin{abstract}
Advances in portability and low cost of plenoptic cameras have revived interest in light field imaging. Light-field imaging has evolved into a technology that enables us to capture richer visual information. This high-dimensional representation of visual data provides a powerful way to understand the scene, with remarkable improvement in traditional computer vision problems such as depth sensing , post-capture refocusing , material classification, segmentation, and video stabilization. Capturing light fields with high spatial-angular resolution and capturing light field video at high frame rates remains a major challenge due to the limited resolution of the sensors, with limited processing speed. In this paper, we presented an extensive literature review of light field acquisition techniques, challenges associated with different capturing methodology and algorithms proposed for light-field super-resolution, in order to deal with spatial-angular resolution trade-off issue.
\end{abstract}


\section{Introduction}
\label{sec:introduction}
Light field imaging techniques can detect rays from different locations as well as different directions. The redundant spatial-angular information recorded in the light-field image or plenoptic image enables many new applications such as post-capture refocusing \cite{ng2005light}, stereoscopic display \cite{katayama1995dependent}, single shot depth measurement \cite{peng2018unsupervised}, view synthesis and 3D displays, with the emergence of commercialized portable light-field cameras, such as Raytrix and Lytro \cite{Lytro} \cite{adelson1991plenoptic}.  By rearranging the measured rays of light to where they would have terminated in slightly different synthetic cameras, we can compute sharp images focused at different depths. As the resolution of the image under each microlens increases linearly, the sharpness of the refocused image also increases linearly \cite{ng2005light}. This property allows us to increase the depth-of-field of the camera without narrowing the aperture, resulting in shorter exposures and less noise in the image \cite{ng2005light}. However, besides several advantages of light field imaging, many researchers believe that there is a fundamental trade-off exists between spatial and angular resolution. In portable light field cameras, the micro-lens array is placed between the main lens and sensor plane, which virtually splits the main lens into sub-apertures that compromise the spatial resolution for angular resolution. Therefore, the spatial-angular resolution of light field imaging limits its promotion to practical applications. Hence, light field super-resolution has drawn more and more attention from researchers and many methods have been proposed to use the redundant information in the 4D light field to handle this challenge \cite{cheng2019light}. Hence, this article provides a critical review of state-of-the-art methods proposed to deal with spatial-angular resolution tradeoff. We have done an extensive literature review of light field acquistion techniques, challenges associated with different capturing methodology and algorithms for light field super-resolution in order to deal with spatial-angular resolution trade-off issue, starting from light-field acquisition via challenges associated with different acquisition methods to Light-field super-resolution.

\subsection{What is Light field ??}
The light fields are vector functions that map the geometry of light rays to the characteristics of the plenoptic system.  The concept of light fields was first proposed in \textbf{1936}. Light fields are becoming increasingly important in computer graphics, especially with the fast growth of computing capacity and network bandwidth.
From a geometric optics views, the plenoptic function defines the set of light rays traveling in every direction through every point in 3D space. In order to obtain this function, one must measure the light rays from every possible angle ($\theta$, $\phi$), at every wavelength ($\gamma$), and at every possible time instant (t). Therefore, the plenoptic function is a 7D function defined as $L(x,y,z,\theta,\phi,\gamma,t)$.\newline
Such high-dimensional data are challenging to record and handle in practice. As a result, for practical purposes, the light field model has been simplified twice. The first assumption is that the function being measured is monochromatic and time-invariant. Each light ray's wavelength ($\gamma$) is recorded independently in each color channel. For a dynamic light field, the time sequence $t$ can be captured in many frames. Thus the dimensions of the plenoptic function is reduced from seven to five. The second simplification was achieved by \textbf{Levoy} and \textbf{Hanharan} \cite{levoy1996light}, as well as \textbf{Gortler et al} \cite{gortler1996lumigraph}, who noticed that 5D representation is reduntant and could be reduced to 4D by assuming that the light field was measured in free space.In such instances, the radiance of the light rays remains constant along the straight line, eliminating the need for one dimension of the plenoptic function.The complexity of measuring and reconstructing the plenoptic function is increased by the additional dimension.
There are three critical issues taken care of while parametrizing the 4D light field. These are computational efficiency, control over the set of rays, and uniform sampling of the light field space. As a result of these issues, the most common solution to represent a 4D light field is to parametrize the light rays by their intersection with two planes placed at random positions. The coordinate system is $(u,v)$ on the first plane, and  $(s,t)$ on the second plane. $L(u,v,s,t)$ is an oriented light ray defined in the system that first intersects the $u-v$ plane at coordinate $(u,v)$ and then crosses the $s-t$ plane at coordinate $(s,t)$. This reduces the plenoptic function of a light field from seven to four dimensions and parametrizes it with four coordinates, \textit{i.e.} $(u,v,s,t)$.\newline

\section{Light field Acquisition Methods}
Light field imaging has emerged as an important technique for capturing more detailed images of our world. As compared to traditional photography, which integrates the angular domain with 2D projection of light in the scene, light fields demultiplex the angular information lost in conventional photography by collecting radiance from rays in all directions \cite{Lightfieldtech}. On the one hand, this higher dimensional representation of visual data offers powerful capabilities for scene analysis, including depth sensing, post-capture refocusing, segmentation, video stabilization, and material classification etc \cite{wu2017}. However, the high dimensionality of light fields also presents new challenges in terms of data capture, data compression, content editing, and display. Together, these two factors have made light field image processing increasingly popular in the fields of computer vision, computer graphics, and signal processing.\newline

A traditional camera records a 2D representation of a light field on its sensor plane by combining the light rays that strike each pixel from all directions. In contrast, the light field acquisition device or techniques measure the spread of light rays in a directionally resolved way, eliminating angular integration. On the other hand, sensors can only measure data from two dimensions (typically two spatial dimensions) of a scene at a time. We need to take several samples along the angular dimensions to get a 4D light field. Existing techniques for the light-field collection may be classified into three broad categories: multi-sensor capture, time-sequential capture, and multiplexed imaging.
\subsection{\textbf{Multi-Sensor Capture}}
The multi-sensor capture technique requires an array of image sensors spread on a flat or spherical surface to acquire light field samples from diverse views concurrently. The sensors determine the light field's spatial dimensions ($u$,$v$), while the number of cameras and their distribution establish the angular dimensions ($s$,$t$). As a result, the 4D light field is created by combining the acquired images.\newline
The Stanford Computer Graphics Laboratory has developed a CMOS-based 100-camera array. The optics, physical spacing, and positioning of the cameras may be altered. Each camera consists of two customized boards: one with a VGA-resolution Omnivision CMOS sensor and a low-cost lens, and the other with a Motorola Coldfire CPU, a Sony MPEG2 video encoder, a Xinilix Field Programmable Gate Array (FPGA), and a Texas Instruments IEEE1394(Firewire)chipset. The system is designed to return live, synchronised, and compressed (8:1 MPEG) video from all 100 cameras at the same time, as well as record these video streams to a stripped disc array through four PCs.Depending on how the cameras are arranged and aimed, multi-camera systems can work in various ways. When the cameras are close together, the system effectively functions as a single-center-of-projection synthetic camera, which can be configured to provide the extraordinary performance along imaging dimensions such as resolution, signal-to-noise ratio, dynamic range, depth-of-field, frame rate, or spectral sensitivity. When the cameras are set further apart, the system operates as a multiple-center-of-projection camera, and the data it collects is referred to as a light field.\newline
\textbf{Yang et al} \cite{yang2002real} described a concept for dynamic light field recording utilizing an array of 8 × 8 video cameras in 2002. They used a distributed rendering algorithm to overcome data bandwidth issues. \textbf{Wilburn et al} \cite{wilburn2001light}  used 6 CMOS image sensors to record synchronized video data for the same year. Each camera included a processing board that used MPEG2 compression to facilitate scalable capture. They increased the system to 125 video cameras and utilized it to collect videos at thousands of frames per second. They investigated the system’s possibilities that would be economical to create in the future. They employed low cost cameras, lenses, and mountings that could process multi-view video. 
 \textbf{Zhang} and \textbf{Chen} \cite{zhang2004self} suggested a self-reconfigurable camera array system that takes video sequences from a mobile camera array produces unique perspectives and reconfigures camera placements to improve rendering quality. The system is made up of 48 cameras that are installed on a mobile platform. \textbf{Chan et al} \cite{chan2005plenoptic} proposed the “plenoptic video,” a method for capturing and producing a dynamic image-based representation. It is a streamlined light field for dynamic situations where user perspectives are limited to the camera plane of a linear array of video cameras. The system comprises an eight-camera array of Sony CCX-Z11 CCD cameras and eight Pentium 4 1,8 GHz processors linked by a 100 Base-T local area network. The system broadcasts plenoptic video (256 x 256 spatial resolution) at 15 frames per second via the network. Liu et al \cite{liu2006real} also created an 8 x 8 dynamic light field streaming system over a broadband network.\newline
 A typical camera array system is large and costly, making it unsuitable for the majority of commercial applications. ViewPLUS Inc.'s ProFusion 25 is a tiny box that houses a number of 5 × 5 VGA cameras \cite{VIEWPLUS}. It can capture light field footage at a rate of 25 frames per second. \textbf{Venkataraman et al} \cite{venkataraman2013picam} presented PiCam (Pelican Imaging Camera-Array), an ultra-thin high-performance monolithic camera array capable of capturing light fields and generating high-resolution pictures (scene depth) via integrated parallax detection and super-resolution. It features a passive camera that can capture both stills and video, is capable of low-light photography, and is tiny enough to fit in the next generation of mobile devices, such as smartphones.\textbf{Pelican Imaging} has created a mobile array camera that can record 3D video and static scenes. This technology has the potential to achieve some very incredible things. Consider how useful it would be if you could refocus after recording a scene, quantify the distance between two spots, or produce a depth map/3D model of the scene.\newline
 \textbf{Lin et al} \cite{lin2015camera} proposed an approach for high-resolution light field microscopy imaging on smaller scene scales by using a camera array. This method utilizes a two-stage relay system to extend the aperture plane of the microscope into the size of an imaging lens array. They use a sensor array to acquire different sub-aperture images formed by the corresponding imaging lenses. This prototype can produce light field videos for various fast-moving microscope specimens with a spatial resolution of 0.79 megapixels at 30 frames per second, corresponding to an unprecedented data throughput of 562.5 MB/sec for light field microscopy. Therefore, multisensor approach can capture the light field instantly, making it capable of acquiring a sequence of light field data. Initially, multisensor systems were bulky and expensive because there were so many cameras.
 \subsection{\textbf{Time-Sequential Capture}}
 Contrary to  multisensor approach, time-sequential capture utilizes a single image sensor to acquire several light field samples over multiple exposures. An optical sensor mounted on a mechanical gantry is often used to measure light fields at various locations. The work describes a sampling representation for light fields that enables the efficient production and display of both inward-looking and outward-looking perspectives \cite{levoy1996light}. From digitized and rendered images, light field can be reconstructed. Additionally, additional views can be created in real-time by extracting slices in the proper direction after generating the light field. The Computer Graphics Laboratory presented two gantry systems at Stanford University \cite{Stanford(New)}: one was a computer-controlled gantry with four degrees of freedom, translation in X and Y, nod, and shake; another was a Lego Mindstorms gantry in which the motors have rotatory controllers enabling the camera to move along accurate and well-defined paths \textbf{Unger et al}. The work \cite{unger2003capturing} presented a process for capturing spatially and directionally varying illumination from a real-world scene and using this lighting to illuminate computer-generated objects. They used two devices for capturing such illumination. In the first, they photographed an array of mirrored spheres in a high dynamic range to capture the spatially varying illumination. In the second, they obtained higher resolution data by capturing images with a high dynamic range omnidirectional camera as it traverses across a plane. They applied the light field technique to extrapolate incident illumination to a volume for both methods. They rendered computer-generated objects as illuminated by this captured illumination using a custom shader within an existing global illumination rendering system.\newline
 In a similar design, \textbf{Kim et al} \cite{kim2013scene} reconstructed complex, detailed environments from 3D light fields with only one degree of freedom. Through breaking with a number of established practices in image-based reconstruction, they proposed an algorithm that takes advantage of the coherence in massive light fields. By operating on individual light rays, rather than image patches, the algorithm computes reliable depth estimates around object boundaries. Interior regions that are more homogeneous are processed in a fine-to-coarse manner rather than using standard coarse-to-fine approach. No global optimization is performed in the proposed method. In less detailed areas, the algorithm was able to maintain precise contours with smooth reconstructions. They also developed a sparse representation and a propagation scheme for accurate depth estimation, making this algorithm particularly useful for 3D input and enabling fast and memory-efficient processing of "Gigaray Light fields" on a standard GPU.\newline
 \textbf{Dansereau et al} \cite{dansereau2017wide} created a small optical design by combining a monocentric lens with microlens arrays. Light field (LF) capture and processing give rich textural information and simplify challenging operations in a growing number of computer vision applications. While light field cameras are commercially available, none of them provide a wide field of vision. As a result, they devised a revolutionary method for linking spherical and plane lenses, thereby replacing costly and cumbersome fiber bundles. They built a prototype single-sensor LF camera that rotated the sensor relative to a fixed main lens to imitate a wide-FOV multi-sensor scenario. Therefore, the approaches discussed so far capture images at different viewpoints by moving the image sensor, which requires high precision control and is time-consuming. Fortunately, fast time-sequential capture approaches have also been developed.\newline
 Like \textbf{Taguchi et al} \cite{taguchi2010axial} proposed a system involving a mirror ball and a single camera. A wide field of view (FOV) catadioptric imaging system has been developed using mirrors. Caustics has been highly effective in understanding the nonlinear relationship between incoming and reflected light rays. These maps were analyzed using two-plane light field parameterization, which provided valuable insight into the geometric structure of reflected rays. They investigated how to generate a single-viewpoint virtual perspective image for catadioptric systems, which is impossible in several common configurations. Instead of minimizing distortion in a single image, they proposed capturing all rays required to achieve virtual perspective by capturing the light field. They considered rotationally symmetric mirrors and showed that a traditional planar light field produced significant aliasing artifacts. In this way, wide FOV virtual perspectives are generated computationally, using a wider range of mirrors than before, without relying on scene priors or depth estimation. Using the camera, they captured an "\textbf{axial light field}" by moving it along the mirror's axis of rotation.\newline
 \textbf{Liang et al} \cite{liang2008programmable} used a programmable aperture instead of a reflective surface to acquire light fields. They presented a novel system incorporating a programmable aperture and two associated post-processing algorithms for high-quality light field acquisitions. An opaque slit of paper or a liquid crystal array encoded the aperture patterns, allowing the camera to capture light rays from certain angles. The programmable aperture's shape can be modified and used to capture a light field at full sensor resolution over several exposures without the need for extra optics or relocating the camera. The complete capture technique required 25 patterns for the light field and each exposure took 10 to 20ms. Each image has a spatial resolution of 3039 x 2014, but it was downsampled to 640 x 426 to ensure excellent processing efficiency. As a result, when compared to multi-sensor systems, time-sequential capture systems require only a single sensor, reducing overall system cost. Moreover, time-sequential capture systems may collect light fields with dense angular resolutions, something that multi-sensor systems cannot do due to the high associated cost. However, because the capture method in time-sequential is generally time-consuming, they are only suited for static scenes. \newline
 \subsection{\textbf{Multiplexed Imaging}}
 The last strategy seeks to encode the four-dimensional light field into the two-dimensional sensor plane by multiplexing the angular domain into the spatial (or frequency) domain. This enables a single image sensor to capture dynamic light fields, but it comes at the expense of spatial and angular resolution tradeoff. A densely sampled image in the spatial region and a sparse sample in the angular region can be obtained. The opposite is also conceivable. Multiplex imaging can be divided into two types: spatial multiplexing and frequency multiplexing.\newline
 \textbf{Spatial Multiplexing}: In spatial multiplexing, an interlaced array of elemental images representing samples from different 2D slices of the light field are captured by the sensor. Most spatial multiplexing approaches are implemented using an array of microlenses or a lenslet array mounted on the image sensor. Ordinary cameras gather light across the area of their lens aperture. The light striking the given sub-region of an aperture is structured somewhat differently from the light striking an adjacent sub-region. By analyzing this optical structure, one can infer the depths of objects from the scene in the scene, \textit{i.e.} one can achieve “single lens stereo.” Therefore, in 1992, \textbf{Adelson and Wang} \cite{adelson1992single} described a novel camera for performing this analysis. It contains a single main lens with a lenticular array placed in the sensor plane. The resulting "plenoptic camera" provides information about what the scene looks like when viewed from a continuum of viewpoints that can be limited by the aperture of the main lens. Derivation of depth information is easier than with a binocular stereo system because correspondence issues are minimized.The device can capture the light field at a 5x5 angular resolution  with 100 x 100 pixels for each image.\newline
 \textbf{Ng et al} \cite{ng2005light} presented a camera that scans the 4D light field on the sensor in a single photoshoot. This is achieved by inserting a microlens array between the sensor and the main lens to create a plenoptic camera. Each microlens measures the total amount of light emitted at that time and the amount of light that travels along each ray. Sharp images focused on different depths by redirecting the measured rays to the final location with a slightly different synthetic camera can be calculated. They showed that as the image resolution under each microlens increased linearly, the sharpness of the refocused image increased linearly. This property allows increasing the camera’s depth-of-field without narrowing the aperture, resulting in shorter exposures and less noise in the image. So, they introduced a handheld light field camera using a 296x296 lens array between the sensor and the main lens. For photographers, plenoptic cameras work just like regular handheld cameras. They used prototypes to take hundreds of light-field photos and showed examples of portraits, fast action, and macro close-ups.\newline
 The above spatial multiplexing device is of the type known as "\textbf{Plenoptic Camera 1.0}". Each microlens on the device captures a position in the scene. There are two types of plenoptic 1.0 cameras on the market, \textbf{Raytrix} \cite{Lytro}, and \textbf{Lytro} \cite{Lytro}. These are for industrial or consumer use. Traditional cameras do not measure sample irregularities or lens aberrations. They have shown that proper use of such irregularities and aberrations can improve light field cameras' quality and ease of use. Traditional cameras and light field cameras are fundamentally different. Optimization of a single captured 2D image and a series of reprojected 2D images from the captured 4D light field. \textbf{Wei et al} \cite{wei2015improving} designs of lens aberrations and microlens / optical sensor sample patterns were evaluated by simulated measurements, and the results were captured by a hardware prototype.Light field detection approaches also use filters and mirror arrays instead of lenslet arrays. \textbf{Horstmeyer et al}  \cite{horstmeyer2009flexible} developed a conventional camera to collect multimodal images with only one exposure. Create synthetic images containing specific spectra, polarization measurements, and optically filtered data. This is done by combining a light field architecture with multiple filters applied to the main lens' pupil surface. The ease with which these filters can be replaced and reconfigured provides excellent flexibility in the type of information that can be extracted from each image.\newline
 \textbf{Manakov et al} \cite{manakov2013reconfigurable} also proposed a non-permanent add-on that enables plenoptic imaging. According to their design, the sensor image is multiplied by multiple identical copies, and the plenoptic data is stored in those copies. It is possible to restore information using a variety of optical filters. With minor design changes, aperture sub-sampling and light-field imaging are also possible. Filters are interchangeable and can be reconfigured for a variety of imaging purposes. Spatial multiplexing has been the most extensively used method for capturing light fields. It is feasible to capture the entire light field in a single photograph with a single exposure. The compromise between angular and spatial resolution at the image sensor, however, is a challenge inherent in spatial multiplexing techniques. \textbf{Georgiev et al} \cite{georgiev2006spatio} solved this challenge by sacrificing angular resolution for higher spatial resolutions in light fields. They  have created a prototype integral camera that attaches to a conventional camera via a system of lenses and prisms. This prototype was used to record the light fields of various sceneries.\newline
  \textbf{Frequency Multiplexing}: In contrast to the spatial multiplexing approach, which interleaves 2D light field slices at the sensor level, the frequency multiplexing approach encodes different 2D light field slices into different frequency bands. Frequency division multiplexing approaches typically use modulation masks to achieve specific characteristics in the Fourier region \cite{ihrke2010theory}.\newline
  
  \textbf{Veeraraghvan et al} \cite{veeraraghavan2007dappled} described theoretical framework for reversible modulation of a 4D light field using an attenuation mask in the optical path of a lens-based camera. Based on this framework, they presented a new design for reconstructing a 4D light field from a 2D camera image without the additional refraction elements required by previous light field cameras. The patterned mask attenuates the rays in the camera rather than bending them, and the attenuation reversibly encodes the rays on the 2D sensor. The masked camera focuses like a traditional camera and captures traditional 2D photos at full sensor resolution, but the raw pixel values also include a modulated 4D light field.They can be restored by tiling 2D Fourier transform of the sensor values into the 4D plane and calculating the inverse Fourier transform.\newline
   \textbf{Lanman et al} \cite{lanman2008shield} explained the unified representation of occluders in optical transmission and shielded field photography. A 4D light-field function that acts on the light field incident on the occluder. The main theoretical result is that \textbf{shields can be used to separate the effects of occluder and incident light}. Lightfield imagers such as Plenoptic and Integral basically compromise the projection of a four-dimensional (4D) light-field scalar function onto a two-dimensional sensor between spatial and angular resolution. A recently proposed programmable light field imager overcomes this spatial angular resolution trade-off, resulting in the high-resolution acquisition of  (4D) light field functions in multiple measurements at the expense of the long exposure time. However, these light field imagers do not take advantage of the angular spatial correlation inherent in the light field in natural scenes, resulting in poor measurement of light efficiency. They described two architectures of compressed light-field imaging that require relatively low photon efficiency measurements to obtain high-resolution estimates of the light field while reducing the overall exposure time.\newline
   
   Using learning-based techniques, \textbf{Marwah et al} \cite{marwah2013compressive} proposed a compressed light field camera architecture that can reconstruct the light field from a single image with a higher resolution than before. This architecture contains three main components \textit{i.e.} a Light-field atom as a sparse representation of a natural light field, an optical design that allows the acquisition of optimized 2D light field projections, and a robust sparse reconstruction to restore a single encoded 2D to 4D light field. Frequency and spatial multiplexing are two different multiplexing imaging approaches, but in reality they are closely related in terms of multiplexing patterns.\newline
   Thus, based on the different approaches discussed so far, it can be concluded that most multisensor approaches can capture light field video but require many cameras. In contrast, time-sequential approaches use a single camera. Despite their ability to capture high spatial and angular resolution light fields, they are impractical for dynamic scenes due to their lengthy capturing procedure. A single-camera can capture high-speed light fields via multiplexed imaging, but there is a tradeoff in angular and spatial resolutions. At the commercial level, light field acquisition devices use spatial multiplexed imaging techniques. Due to a large amount of data that needs to be recorded, however, light field video recording is difficult. Lytro Illum, for example, can only record light field video at 3fps. By using frequency division multiplexing, modulation masks lose luminous flux, resulting in longer exposure times. The miniaturization of sensors enables integration into smaller monolithic camera arrays or the installation of multiple sensors in a portable box. As a result of this trend, light field cameras may become common in mobile devices such as mobile phones, tablets, and computers in the near future.\newline
   
\section{Light-Field Super-resolution methods}
Light-field super-resolution methods are boardly classified into following types \textit{i.e.} Spatial super-resolution, Angular super-resolution, Temporal super-resolution and Spectral reconstruction. 
\subsection{\textbf{Spatial Super-resolution }} 
The light field captured by Lytro Illum has a resolution of 7728x5368 pixels, which is larger than most DSLR cameras today. However, the resolution of a single sub-aperture image extracted with Light Field Matlab Toolbox\cite{Matlab} is only 625x434. An easy way to reconstruct a high resolution light field in spatial dimensions is to transform it using a single image super-resolution method.  However, this type of method does not maximize the potential information present in the light field because each image in the light field is treated as a separate entity.
As explained in \cite{georgiev2009superresolution}, the digital implementation of \textbf{Lippmann's integrated photography} \cite{lippmann1908epreuves}, the plenoptic camera,  was introduced in 1992 as an approach to solving computer vision problems. An improved version is the Plenoptic 2.0 camera was introduced independently \cite{georgiev2009superresolution}. The Plenoptic 2.0 camera also follows the idea of lippmann \cite{lippmann1908epreuves}. Super-resolution is based on extracting sub-pixel information from multiple images in a particular scene to create a high-resolution image. The work  showed that the Plenoptic 2.0 camera can be interpreted as a series of cameras  focused on objects taken through  relay optics \cite{georgiev2009superresolution}. Based on this interpretation, they have developed super-resolution technology for rendering high-resolution images from the acquired data. The geometry of Microlens Array allows super-resolution to be applied without traditional registration in the software.\newline
Another method has developed a light transport framework to understand the basic resolution limitations of light field cameras \cite{liang2015light}. They first derived a pre-filtering model for a lenslet-based light field camera. The main novelty of the model  is to consider the complete space angle sensitivity profile of the optical sensor \cite{liang2015light}. Especially for real pixels that have non-uniform angular sensitivity and react to light along the optical axis . The work has shown that a complete sensor profile plays an important role in defining the performance of a light field camera \cite{liang2015light}. This method  can model all existing lenslet-based light field cameras and compare them in a simulation-consistent manner, regardless of the actual differences between the individual models \cite{liang2015light}. They further extended their framework to analyze the performance of two rendering methods \textit{i.e.} the simple projection-based method and the inverse light transport process. In a 4D light field captured by a plenoptic camera, each image is a 2D sample from one viewpoint, the difference from adjacent views is small, and a non-integer shift between the two corresponding pixels in the image.  By taking advantage of these non-integer shifts,the pixels can be propogated from the  adjacent views to the target view.\newline

Following the general principle above, a typical \textbf{spatial super-resolution method} first estimates the depth of the scene to infer these non-integer shifts and then uses a different optimization framework to image for super-resolution.\newline

\textbf{Chan et al} \cite{chan2007super} explicitly introduced the priority of Lambertian reflection into the imaging model and proposed a mathematical model that simulates the super-resolution of the light field. From household appliances to biomedical applications, device miniaturization has proven highly desirable. This often involves reducing the size of some optical systems. However, if we simply reduce the image parameters, the diffraction effect will limit the image quality. In recent years,  Compound eye imaging systems have emerged as a promising architecture in the development of compact visual systems. Since multiple low resolution (LR) frames are acquired, the post-processing algorithm for reconstructing the final high resolution (HR) image from the LR image plays an important role in affecting  image quality. This work explained and investigated the performance of compound eye systems recently reported in the literature \cite{chan2007super}. Low resolution and low image quality are the main challenges of Compound eye imaging systems. They published an analysis of a compound eye system similar to the \textbf{TOMB} (Thin observation module by bound optics) architecture. They described both  physical structure and  mathematical modeling, along with their own super-resolution image reconstruction algorithms, to improve the quality and resolution of the re-conceived images. They also considered  extending the basic Compound eye system, such as: 
 Include a phase mask in some of the paths to take advantage of the additional flexibility provided by different paths.\newline
 
 \textbf{Bishop et al} \cite{bishop2011light} have shown that they have an extended depth-of-field, but retain the ability to reconstruct high-resolution details. In fact, all depths are almost in focus, except for one thin slab, which has a limited blur size. Their depth-of-field is essentially inverted compared to regular cameras. Key to their success is the LF sampling method, the spatial and angular resolution trade-offs, and how aliasing affects  LF. The work has shown that applying traditional multi-view stereo methods to extracted low-resolution views can result in reconstruction errors due to aliasing \cite{bishop2011light}. They addressed these challenges with an explicit image generation model, integrated Lambertian prior distributions with texture, and reconstructed the scene depth and super-resolved textures in variational Bayesian frameworks. However, aliasing has been eliminated by fusing multi-view information. This work showed how synthetic images and real images were captured by light field cameras and showed superiority over other computational camera systems. The contributions by the method proposed are as follows: 1. An explicit imaging model was obtained by characterizing the spatially variable PSF of the plenoptic camera, assuming Gaussian optics at various depths of scene. 2. Novel analysis of view aliasing  and plenoptic camera DoF (taking into account non-negligible blur created by sampling and  microlenses) 3. How to reconstruct a light field in  Bayesian framework. Explicitly introduce Lambertian reflection priors into the imaging model. Note that this allows designing of super-resolution algorithms that recover more information than was predicted by the basic sampling theorem 4. A method of reducing view aliasing by spatially varying the filtering of  recorded LF, and an iterative multi-view depth estimation method that benefits from this reduction; 5. A comparison of resolutions that are actually achievable at different depths compared to other methods using SR approach. This work demonstrated that  LF cameras are superior to regular, coded aperture cameras and focus sweep cameras in noisy situations.\newline

\textbf{Zhou et al} \cite{zhou2017multiframe} presented how to reconstruct a high-resolution image from an angular image. First, they used the ray tracing method  to model the telecentric light field imaging process. Next, they analyzed the subpixel shift between the angular image extracted from the defocused light field data and blur of the angular image. By applying the regularized super-resolution method, They obtained a super-resolution result with a magnification  of 8. Therefore, they modeled telecentric light field imaging process and reconstructed the observation model between ideal HR  and angular images. In the observation model, the observation matrix can be obtained analytically. The observation model uses a regularized super-resolution method  to reconstruct the HR image.\newline

Lightfield photography opens up many new possibilities for digital imaging because it captures both spatial and angular information, the complete four-dimensional information of the scene. To be able to capture 4D data with a 2D sensor requires a very high resolution. However, the image rendered from  light field as a projection of 4D radiation onto two spatial dimensions has a much lower resolution. \textbf{Georgiev et al} \cite{lumsdaine2008full} in order to meet the expectations of modern digital photography resolution and image size, proposed new technology for reproducing high-resolution images from the light field. This approach is called full resolution because it makes the best use of both the location and angular information available in the acquired radiation data. They have experimentally demonstrated the effectiveness of this method by rendering images from a 542-megapixel light field using the traditional and proposed approach.\newline

\textbf{Reference-Based Super-Resolution} (RefSR) super-resolves low-resolution (LR) images given an external high-resolution (HR) reference image.The reference image and LR image share the same angle of view, but with a large resolution gap. Existing RefSR methods work in a cascade way, \textit{i.e.}, a synthetic pipeline with patch matching followed by  two independently defined objective functions result in inconsistencies, grid effects, and inefficient optimization between patches. To solve these problems, \textbf{Zheng et al} \cite{zheng2018crossnet} introduced \textbf{CrossNet}, a fully convolutional end-to-end deep neural network with cross-scale warping. This network contains an image encoder and a crossscale warping layer that spatially aligns reference feature maps from both domains to synthesize  HR output. Cross-scale warping allows networks to perform pixel-level spatial alignment end-to-end, improving existing schemes in terms of both accuracy and efficiency.\newline

\textbf{Mitra and Veeraraghavan} \cite{mitra2012light} provided a common framework for many light field processing tasks such as denoising, angular and spatial super-resolution by designing the patch prior from the "disparity pattern" of the light field data. In this work, light field patches with the same disparity value (\textit{i.e.} the same depth from the focal plane) on a low dimensional subspace have been considered, and the dimensions of such a subspace change quadratically with the disparity value \cite{mitra2012light}. Next, they modeled the patch as a Gaussian random variable that depends on the disparity value and effectively created the GMM model. Therefore, the technical contributions of this method are: 1. In this work, a common framework has been proposed for solving many different lightfield processing tasks using  GMM prior to the lightfield patch. 2. In this work, patches with a common disparity value were in a lower-dimensional subspace. They calculated the dimensions of these subspaces for different disparity values. This provides an estimate of the minimum number of observations needed to reliably build light-field patches with different disparity values. 3. In this work an efficient algorithm has been used for solving the inference problem.  Fast subspace projection algorithm has been used to estimate the observed patch disparity values and then  LMMSE(Linear minimum mean squared error) algorithm used to compute the final result.\newline

\textbf{Alain et al} \cite{alain2018light} proposed a light field spatial super-resolution method that combines  \textbf{SRBM3D} single-image super-resolution filter with  recently introduced \textbf{LFBM5D} light-field noise reduction filter. The proposed algorithm iteratively alternates between  LFBM5D filter step and back-projection step. The LFBM5D filter produces a disparity-corrected 4D patch. These patches  are  stacked along the 5th dimension with similar 4D patches. The 5D patch is then filtered by 5D transform domain and sparse-coded in the high resolution light field. This is powerful before solving ill-posed super-resolution problems. The back projection step then forces a match between known low resolution light field and the high resolution estimate.This step further improved by removing ringing artifacts using  image-guided filtering. Therefore in this work \textbf{iterative SRBM3D single-image super-resolution method} , which is derived from  \textbf{BM3D single-image noise reduction filter} combined with  \textbf{LFBM5D light-field noise reduction filter}, which extends the concept of \textbf{BM3D} to light field. First, the \textbf{LFBM5D filter} is applied to current estimates of High-Resolution LF. In this step, the 5D patch is first converted to 5D transform domain. Due to high redundancy of light field patches, the resulting 5D spectrum is very sparse and coefficients are also hard-thresholded. This step can be interpreted as forcing a sparse priority on the light field. The back-projection is then applied to each sub-aperture image (SAI). Back-projection consists of up-sampling the residual error between a known light field image and the current down-sampled HR estimate. The up-sampling residual error is then added  to the current HR estimate to create a new estimate. The process repeats until it converges. Back-projection is affected by known ringing artifacts, especially at high magnification. Therefore, image-guided filtering is used to improve this step. \newline

Previous attempts aimed at improving the spatial resolution of plenoptic light-field images are \textbf{block} and \textbf{patch matching} inherited from classic image super-resolution methods where multiple views are taken as separate frames. In contrast to these approaches, \textbf{Farag et al} \cite{farag2018novel} proposed a new super-resolution approach. It focuses on reducing the matching area in super-resolution process using the estimated disparity information. This method estimated disparity information from the interpolated LR viewpoint image. This method called as \textbf{light field block matching super-resolution}.\newline

\textbf{Rossi and Frossard} \cite{rossi2017graph} in  \textbf{2017} proposed a new lightfield super-resolution algorithm that provides a global solution that increases the resolution of all views together without relying on explicit  disparity estimation steps or offline learning procedures. In particular,  spatial super-resolution of the light field has been transformed into a global optimization problem with an objective function with three terms. First, it enhances data fidelity by constraining each high-resolution view to match the low-resolution view. The second is a warping term that collects complementary information encoded in other views for each view. The third is a graph-based prior that adjusts the high-resolution view by applying a geometric lightfield structure. Together, these terms form a quadratic objective function that can be solved iteratively. \newline

\textbf{Rossi and Frossard} \cite{rossi2018nonsmooth} again proposed a new super-resolution algorithm in \textbf{2018}. There, it turns lightfield super-resolution into an optimization problem, where certain structures of lightfield data are captured by a \textbf{non-smooth graph-based regularizer}. All light field views are commonly super-resolved. Therefore, they have developed a super-resolution algorithm that raises the resolution of all  views together, without an offline learning procedure, relying only on a very rough estimate of the disparity in each view. In particular, the light field super-resolution applies to global optimization problems where the objective function consists of three terms. First, it enhances data fidelity by constraining each high-resolution view to match the low-resolution view. The second is a warping term that collects complementary information encoded in other views for each view. The third is a novel graph-based prior that adjusts the high resolution view by enforcing the geometry of the light field. Unlike quadratic graph-based regularization, this graph-based regularization is not smooth. This represents an important difference, as secondary regularizers are known to guide low-pass filtered solutions. In general, textures, edges, and fine structures are better preserved.\newline

\textbf{Wanner et al} \cite{wanner2013variational} developed a continuous framework for  4D light field analysis and describe new variational methods for disparity reconstruction and spatial and angular super-resolution. Disparity maps are estimated locally using epipolar planar image analysis without the need for expensive cost matching minimization. This method worked quickly and with inherent subpixel accuracy because it did not require discretization of the disparity space. The variational framework used disparity maps to generate a new super-resolution view of the scene, much like increasing the sampling rate of a 4D light field both spatially and angularly. Unlike previous work, they formulated the view composition problem  as a continuous inverse problem. This can correctly explain shortening effect caused by the geometry transformation of the scene.Therefore,  this method first addressed problem of disparity estimation in the light field and introduced a new local data term for the continuous structure of the light field. The proposed method can obtain locally  robust results very quickly without the need to quantize the disparity space into discrete disparity values. Local results can be further integrated into a globally consistent depth map using state-of-the-art labeling schemes based on convex relaxation techniques \cite{pock2010global}\cite{siu2005image}. In this way, this method obtained accurate geometry estimates with subpixel precision matching. It can be used to simultaneously address  spatial and angular super-resolution problems. To speed up, this method also provided a new fast way to combine disparity estimates from local epipolar plane images (EPIs) into global parallax maps. It takes advantage of the fact that in a light field where the viewpoint is \textbf{densely} sampled, the \textbf{derivative} of the intensity  with respect to the \textbf{position} of the viewpoint can be calculated. The most notable difference from the standard disparity estimation is that it did not actually calculate the  stereo correspondence in a normal scene. As a result, the method execution time is completely independent of the required disparity resolution. \newline

\textbf{Boominathan et al} \cite{boominathan2014improving}, as opposed to the above approaches, used inherent information captured by plenoptic cameras. This method proposed a hybrid imaging system consisting of a standard LF camera and a standard high resolution (HR) camera, which enables:a) Achieving high-resolution digital refocussing. b) Better DOF control than LF cameras. c) Rendering graceful high-resolution viewpoint variations. \newline

All of this was previously unattainable. Therefore, this method  proposed a simple patch-based algorithm that super-resolves a low-resolution (LR) view of the light field using high-resolution patches captured by a high-resolution single-lens reflex camera \cite{boominathan2014improving}. This algorithm does not require the LF camera and  DSLR to be co-located and does not require the presence of calibration information for the two imaging systems. They created an example prototype using a Lytro camera (380 x 380 pixels spatial resolution) and an 18 megapixel (MP) Canon DSLR camera, with 11 MP resolution (9 x super-resolution) and about 1/9th of the depth of field of the lytro camera. The hybrid imager they propose is a combination of two imaging systems, a low-resolution light field device (Lytro camera) and a high-resolution camera (DSLR). The Lytro camera captures the  perspective view and  depth of the scene, and the DSLR camera captures a picture of the scene. The algorithm combines these two imaging systems to produce a light field with DSLR spatial resolution and Lytro angular resolution.\newline

Using this approach, \textbf{Wang et al} \cite{wang2016light} proposed a lens-based concept that turns a standard DSLR camera and lens into a light field camera. The attachment consists of eight low-resolution, low-quality side cameras placed around a central SLR lens. Different from most existing light field camera models, this design provides a high quality 2D imaging mode, enabling a new high-quality light field mode with a large camera baseline. And there is little additional weight, cost, or bulk. From an algorithmic point of view, high-quality light field mode is made possible by the new light field super-resolution method. First, improve the spatial resolution and image quality of the side camera, and further interpolate the views as needed. At the heart of this process is a super-resolution technology called \textbf{Iterative Patch and Depth-Based Composition} \textbf{(iPADS)} that combines patch-based and depth-based composition in a novel way.\newline

In particular, they proposed a method called \textbf{iterative Patch and Depth Based Synthesis} (iPADS) for light-field super-resolution using low-resolution side views combined with a central high-resolution digital SLR image. An important idea for iPADS is to provide high-resolution patch candidates that are more similar to the high-resolution ground truth of the side view image compared to the available center image. By leveraging depth information using a phase-based rendering approach, these patches are rendered from the center image and retain the high-frequency details of the center image. Therefore, the main technical contributions of this method are:\newline

1. This method presented the concept of a light field lens attachment that can transform a DSLR camera into a  high spatial and angular resolution light field camera and developed the first prototype to confirm its feasibility.\newline

2. This method presented an optimization framework called Patch And Depth-based iterative synthesis that used the data captured by the light field attachment to achieve light field super-resolution. The proposed method iterates patch-based and depth-based compositing for super-resolution to provide better patch candidates for achieving light field reconstruction with high spatial and angular resolution.\newline

3. This method proposed a new depth-based synthesis method that can co-synthesize high-resolution side views using high-quality center view textures and estimate center view depth with high quality.\newline

\textbf{Yoon et al} \cite{yoon2015learning} used a convolutional neural network (CNN) and presented a new method of light field super-resolution (SR) imaging via deep convolutional neural networks. Instead of the traditional optimization framework,  used a data-driven, learning-based method to simultaneously estimate the angular and spatial resolution of a light field image with no depth estimation step. In this method, first increased the spatial resolution of each sub-aperture image to emphasize the details via the spatial SR network. Next, a new view is generated between the sub-aperture images via the angular super-resolution network. These networks are trained independently but ultimately refined through end-to-end training.Therefore, a deep convolutional network has been introduced for super-resolution of light-field images \cite{yoon2015learning} . This network consists of the spatial SR network and angular SR network to improve the spatial and angular resolution comprehensively. Therefore, this method can generate a high-resolution sub-aperture image in the novel view between adjacent sub-aperture views.\newline

\textbf{Farrugia et al} \cite{farrugia2017super} described a light-field example-based super-resolution algorithm that allowed us to increase the spatial resolution of different views in a consistent manner across all sub-aperture images of the light field. The algorithm learned a linear projection between reduced dimensional subspaces containing patch volumes extracted from the light field. This method has been extended to use multivariate ridge regression to handle angular super-resolution in which a 2D patch of an intermediate sub-aperture image is approximated from an adjacent sub-aperture image. Therefore, this method works with a 3D stack (called a patch volume) of 2D patches extracted from different sub-aperture images to maintain consistency across all sub-aperture images in the light field, an example-based spatial super-resolution. The patches that make up 3D stack are either co-localized patches or the patches that best match across the sub-aperture image.First, an example dictionary  is created from a high resolution and low resolution light field training set, a pair of high resolution and low resolution patch volumes. These patch volumes are  very high dimensional $ q $ x $ q $ x $ n $. Where $ n $ is the number of sub-aperture images and $ q $ x $ q $ is the size of each 2D patch. Nevertheless, these patch volumes are actually in low-dimensional subspaces because  of redundant information. Therefore, each pair of low-resolution and high-resolution patch volumes can be projected into their respective low-resolution and high-resolution subspaces. Below, the projection \textbf {principal component analysis} is performed. A dictionary of projected patch volume pairs (example) locally maps relationship between high-resolution patch volumes and those low-resolution patch volumes. Then use \textbf {Multivariate Ridge Regression} (RR) to learn a linear mapping function  between  subspaces of  low and high resolution patch volumes. Then learned mapping function is applied directly to super-resolution for each overlapping patch volume of the low-resolution light field.\newline

\subsection{\textbf{Angular Super-resolution}}
Much research has focused on angular super-resolution using a small set of high spatial resolution views. These views can be modeled as a reconstruction of plenoptic functionality with limited samples. Existing angular super-resolution approaches can be divided into two categories \textit{i.e.} \textbf{Depth-Based} and \textbf{Without Depth-Based}.\newline

 \subsubsection{\textbf{Depth Image-Based View synthesis}}
 The approach of synthesizing a view based on a depth image usually first estimates depth information and then warps existing image to new view, based on estimated depth.Warped views are blended in a specific way.\newline

\textbf{Todor Georgeiv} \cite{georgiev2006spatio} has developed a prototype of an integrated camera that uses a lens and prism system as an external attachment to a traditional camera. This method  used this prototype to capture the light fields of different scenes \cite{georgiev2006spatio}. In this method a segmentation-based optical flow method is used to calculate the flow between an image and two adjacent views. Then, three warped views have been weighted to create a new one.\newline

Image-based rendering is less complex and can give photo realistic results, so it can able to generate novel views in a compelling alternative to model-based rendering. To reduce the number of images required for alias-free rendering, usually need some geometric information in the 3D scene. Therefore, \textbf{Pearson et al}. \cite{pearson2013plenoptic} presented a fast, automatic layer-based method for synthesizing any new view of the scene from an existing set of views. Their  algorithm uses knowledge of the typical structure of multi-view data to perform occlusion-aware layer extraction. Also, the \textbf{number of depth layers used to approximate the geometry of scene is selected based on  plenoptic sampling theory, and the layers are evenly spaced to take into account distribution of the scene}. Rendering is achieved by using a probabilistic interpolation approach and extracting depth layer information from a small number of key images.\newline

\textbf{Wanner et al}. \cite{wanner2012spatial}\cite{wanner2013variational} formulated the view synthesis problem as an energy minimization problem.\newline

Advanced camera calibration and multi-view stereo technology allow users to smoothly navigate between different views of a scene captured by a standard camera. The underlying  3D automated reconstruction method is suitable for buildings and regular structures,but often fails for vegetation, vehicles, and other complex geometries found in everyday urban scenes. As a result, the lack of depth information makes image-based rendering (IBR) of such scenes a major challenge. Therefore, the goal is to provide plausible Free viewpoint navigation for such datasets. \newline
Therefore \textbf{Chaurasia et al} \cite{chaurasia2013depth} introduced a new Image-Based-Rendering algorithm that is robust against missing and unreliable geometry and provides a plausible new view even in areas well away from the input camera. This method, first over-segmented the input image to create superpixels with uniform color content that tended to maintain depth discontinuities. Next,  introduced a depth synthesis approach for poorly reconstructed areas based on the graph structure for graph over-segmentation and proper traverse. Superpixels enhanced with combined depth allowed us to define a local shape that holds the warp to correct for inaccurate depth. Their  rendering algorithm blends warped images to produce a novel image-based view that is plausible for challenging target scenes       \cite{chaurasia2013depth}.\newline

\textbf{Pujades and Devernay} \cite{pujades2014bayesian} proposed a new physics-based generative model and its corresponding maximum post-estimation, providing the desired coupling between heuristic-based methods and  Bayesian formulation. The important point is to systematically consider the errors caused by the uncertainty of the geometric proxy. Therefore, in this method a view synthesis approach has been proposed by optimizing a new cost function using a robust Bayesian formulation for estimated depth errors.\newline

\textbf{Zhang et al} \cite{zhang2015light} presented a new phase-based approach to reconstruct  4D light fields from micro-baseline stereo pairs. This approach takes advantage of unique characteristics of the complex steerable pyramid filters in micro-baseline stereo pair.In this method  a \textbf{Disparity-assisted phase-based synthesis} (DAPS) strategy is introduced that can integrate disparity information into the phase term of reference image and warp it into adjacent views. Based on  DAPS, an \textbf{"analysis by synthesis"} approach was proposed, warping from one of the input binocular images to another, repeatedly optimizing the disparity map.  Finally, we can use the DAPS proposed according to a sophisticated disparity map to reconstruct a high-quality, dense, regularly spaced light-field image. This approach also solves the problems of disparity mismatch and ringing artifacts in the available phase-based view synthesis methods. Therefore, in this method a new phase-based light field synthesis architecture has been proposed that can reconstruct high-quality, high density sampled light fields from micro-baseline stereo pairs. In this work, the micro-baseline refers to a pair of stereo images with a disparity of less than 5 pixels.\newline

The patch-based image composition method has been successfully applied to various editing tasks for still images, video and stereo pairs. \textbf{Zhang et al} \cite{zhang2016plenopatch} extended patch-based synthesis to plenoptic images captured by consumer lenslet-based devices for interactive and efficient light field processing. In this way, the light field is represented as a series of images taken from different perspectives. This method first decomposed the center view into various depth layers and presented them to the user to specify editing goals. For editing tasks, this method performs patch-based rendering on all layers affected by center view and then propagates edits to all other views. The interaction takes place through traditional 2D imaging user interface, which is familiar to novice users. This method is translucent and handles object boundary and occlusion correctly, so it can produce more realistic results than the previous method.\newline

Typical depth-based view synthesis approaches rely heavily on the estimated depth, which is sensitive to textureless and occluded regions. In addition, they often focus on the quality of depth estimation rather than the quality of synthetic views.In recent years, some studies based on CNNs aimed at maximizing the quality of synthetic views have been presented.\newline

\textbf{Flynn et al} \cite{flynn2016deepstereo} introduced a new deep architecture that performs novel view synthesis directly from pixels trained from a large set of posed images. The proposed system is continuously trained, as opposed to the traditional approach, which consists of several complex processing stages, each of which requires careful coordination and can fail in unexpected ways. Pixels from adjacent views of the scene are presented to the network, and the network directly produces hidden view pixels. The advantages of this approach include generality. This method can be easily applied to different areas, resulting in high quality results in traditionally difficult scenes. They believed that this was due to end-to-end nature of the system and could generate pixels according to color, depth, and texture priorities automatically learned from training data \cite{flynn2016deepstereo}.Therefore, in this work,a new approach has been presented to synthesize a new view that traces back the input image  directly to the output pixel color using a deep network, given the input image provided. The proposed system can interpolate between views separated by a wide baseline, demonstrating resilience to common failure modes such as scene movement and graceful degradation in the presence of specular reflections. This method made minimal assumptions about the scene to be rendered. The scene is primarily static and must be within a finite depth range. If these requirements are violated, the resulting image will  be elegantly degraded and remain visually plausible. If uncertainty cannot be avoided, this method prefers to blur details, especially with animations that are much more visually pleasing than tearing or repeating results.\newline

\textbf{Kalantari et al} \cite{kalantari2016learning} proposed a new learning-based approach for synthesizing new views from a sparse set of input views. In this method, machine learning techniques are used to reduce the inherent trade-off between angular and spatial resolution. They are built on existing view synthesis techniques and break down the process into disparity and color estimation components. In this method, two sequential convolutional neural networks are used to model these two components and trained both networks simultaneously by minimizing errors between the synthesized image and ground truth image. This method is two orders of magnitude faster than \textbf{Flynn et al} \cite{flynn2016deepstereo} [2016], it takes only 12.3 seconds to synthesize an image from four 541x376 input views. The proposed system could be used to decrease the angular resolution required for current cameras, thereby increasing their spatial resolution. Therefore, the contributions of this method are as follows:\newline

1. This method presented first machine learning approach to view synthesizing on a consumer light field camera. The proposed system consists of a disparity and color estimation component modeled on two consecutive CNNs.\newline

2. The output of first network is disparity, which usually requires ground truth disparities to train that network.\newline

\subsubsection{\textbf{Light Field Reconstruction without Depth}}
Approaches that synthesize views based on depth images include depth estimation that tends to fail in occluded regions as well as glossy or specular regions. Another approach is based on a sampling of plenoptic function and subsequent reconstruction. In light field rendering, if the samples are inadequate, ghosting will occur in the new view. However, it is not practical to collect too many light field samples. Therefore,  to avoid the ghost effect, the maximum disparity between adjacent views should be less than 1 pixel. This is a value that is closely related to the resolution of camera  and the depth of the scene. \textbf{Chai et al and Lin et al}  \cite{chai2000plenoptic}\cite{lin2004geometric} suggested that the closer the contributing pixels are (that is, the less disparity), the sharper the interpolated points will be.\newline

For light fields with sparse sampling, direct interpolation causes ghosting in the rendered view. To mitigate this effect, some studies have considered reconstructing the light field in the Fourier domain. \textbf{Levin and Durand} \cite{levin2010linear} argued that the fundamental difference between the different capture and rendering methods was the difference in prior assumptions about the light field. This method used the previously reported dimensional gaps in 4D light field spectrum to predict new light fields. The new prior distribution is a Gaussian value that primarily assigns a non-zero variance to the 3D subset of entries. Since there is only a low dimensional subset of entries with nonzero variance, the complexity of the acquisition process can be reduced and render 4D light field from 3D measurement sets. Moreover, the prior with Gaussian nature leads to linear and depth invariant reconstruction algorithms. This method used the new prior to render the 4D light field from a 3D focal stack sequence and to interpolate sparse directional samples  \cite{levin2010linear}. Therefore the algorithm reduces to a simple spatially invariant deconvolution that \textbf{doesn`t involve depth estimation}.\newline

Fourier domain sparsity is an important property that enables high-density reconstruction of signals such as 4D light fields from a small set of samples. The sparseness of the natural spectrum is often derived from the continuous arguments, but reconstruction algorithms usually work in the discrete Fourier domain. These algorithms usually assume that sparsity derived from the continuous principle also applies to discrete sampling. When the signal is  sampled in a finite window, convolution  between the  spectrum of the signal occurs through an infinite sinc, destroying much of the sparsity of the continuous domain. Based on this observation, \textbf{Shi et al} \cite{shi2014light} proposed a reconstruction approach  that optimizes the sparsity of continuous Fourier spectrum.The difference between continuous sparsity and discrete sparsity is due to the window effect. Sampling a signal, such as a light field, in a finite window is similar to multiplying that signal by a box function. In the frequency domain, this multiplication is an infinite sinc convolution. Unless the non-zero frequency of the spectrum exactly coincides with the resulting discretization of the frequency domain (and thus the zero intersection of  sinc), this convolution destroys much of the sparseness that exists in the continuous domain. Therefore, in this method an approach to restore the sparsity of the original continuous spectrum based on the  nonlinear gradient descent method has been introduced. In this method, the sparsity in the continuous frequency domain has been optimized by carefully modeling  the projection of the continuous sparse spectrum into the discrete domain, starting with an initial approximation of the spectrum. The output of this process is an approximation of the continuous spectrum. For light fields, we can also use this approximation to reconstruct a high-quality view that has never been captured. The proposed method effectively reduces the sampling requirements of 4D light field by restoring the sparsity of the original continuous spectrum. This method  shows that a complete 4D light field can be reconstructed from  a 1D perspective trajectory. This can greatly simplify the capture of light fields.\newline

Naked-eye stereoscopic multi-view displays provide an immersive, eyeglass-free 3D viewing experience, but require properly filtered content from multiple perspectives. However, this is not easily possible with current stereoscopic production pipeline.\newline

\textbf{Didyke et al} \cite{didyk2013joint} provided a practical solution to take stereoscopic video as input and convert it into a multiview and filtered video stream that can be used to drive an automatic stereoscopic multiview display. This method combines phase-based video magnification with inter perspective antialiasing in a single filtering process. The overall algorithm is simple and can be efficiently implemented on today's GPUs for near real-time performance. This method is robust and suitable for difficult video scenes such as defocus, motion blur, transparent materials, and specularities.Therefore, phase information from a complex steerable pyramid decomposition has been used to synthesize a novel view with small parallax. However, these reconstruction approaches always require the light field to be sampled in a specific pattern. This is a real application limitation.\newline

An image-based reconstruction technique introduced by \textbf{Vagharshakyan et al}  \cite{vagharshakyan2017light}\cite{vagharshakyan2017accelerated} is based on light field reconstruction from a limited set of perspective views captured by a camera. This approach uses a sparse representation of the epipolar plane image (EPI) of the shearlet transform domain. The shearlet transform has been specially modified to handle the straight lines characteristic of EPI. The iterative regularization algorithm developed based on the adaptive threshold provides high-quality reconstruction results for relatively large disparities between adjacent views. Therefore, in this method, the concept of light-field sparsification and depth layering have been advanced with aim of developing an effective reconstruction of light-field represented by EPI. Reconstruction attempts to use the appropriate transformations that provide a sparse representation of EPI. It is assumed that a good sparse transformation should include a scene representation with a depth layer that is expected to be sparse. Since the anisotropic properties of  EPI are caused by shear transformation, this method preferred shearlet transformation as the sparse transformation  and developed impainting techniques that act on EPI. Therefore, this method approached angular super-resolution as an EPI impainting problem and super-resolved the angular resolution of the light field using an adapted discrete shearlet transformation. \newline

In addition, several learning-based methods for depth-free angular super-resolution were presented. \textbf{Yoon et al} \cite{yoon2015learning} proposed a CNN-based approach to generate the middle view using two adjacent views in the vertical (horizontal) angular dimension. However, this method does not  take advantage of all the angular information possibilities and could only achieve a fixed super-resolution.\newline

On the other hand, by taking advantage of the clear texture structure of the epipolar planar image (EPI) of the light field data, \textbf{Wu et al} \cite{wu2017light} modeled the light field reconstruction problem from a sparse set of views as a CNN-based angular detail recovery at EPI. They point out that one of the major challenges in sparse sampling light field reconstruction is the asymmetry of information between spatial and angular domains, where details of the angular domain are corrupted by under-sampling. To balance spatial and angular information, EPI's high-frequency spatial components are removed using EPI blur before being injected into the network. Finally, this method use a non-blind deblur operation to restore the spatial details suppressed by EPI blur. Therefore, in this method a new learning-based framework  is proposed for reconstructing light fields with high angular resolution from sparse samples of views. One of the key insights is that light-field reconstruction can be modeled as learning-based angular detail restoration on the 2D EPI. The special structure of EPI allows us to effectively implement learning-based reconstruction. In comparison with the depth-based view synthesis approach, the proposed method does not require depth estimation.

\subsection{\textbf{Temporal Super-resolution}}Due to the high dimensional data of the light field, current commercial light field cameras are usually unable to record light field video at a satisfactory frame rate. For example, Lytro Illum.  \cite{Lytro} captures a  7728 x 5368-pixel light field in a single shot and can only achieve 3fps in the continuous shooting mode. Most researchers focus on video frame interpolation \cite{meyer2015phase}\cite{mahajan2009moving}. \newline

A method defined by \textbf{Mahajan et al} \cite{mahajan2009moving} is a way for plausible interpolation of images, with an extensive variety of applications like smooth playback of lower frame rate video by temporal up-sampling, smooth view interpolation, and animation of still images. The approach is primarily based intuitive idea, that a given pixel within the interpolated frames, lines out a direction within the source images. Therefore, this method clearly moved and copied pixel gradients from the input images alongside its direction. A key innovation is to permit arbitrary or uneven transition points, wherein the direction of movements from one image to the other. The flexible transition preserves frequency content of originals with out ghosting or blurring and keeps temporal coherence. Perhaps, maximum importantly, the proposed framework makes occlusion handling particularly simple. The transition points allow for matches away from the occluded regions, at an appropriate point along the path.\newline

The standard approach for computing interpolated (intermediate) frames in a video sequence require accurate pixel correspondence between frames example using optical flow .\textbf{Meyer et al} \cite{meyer2015phase}  presented an efficient alternative using recent developments in phase-based methods that represent  the phase shift motion of individual pixels. This concept allows us to generate intermediate images with a simple pixel-by-pixel phase modification without the need for explicit matching estimation. So far, such methods have been limited to the range of motion that can be interpolated, which is basically limited in usability. To reduce these limitations, in this method a new restricted phase shift correction method has been introduced that combines phase information across  levels in a multiscale pyramid. In this method, an extension of phase-based image synthesis that provides smoother transitions between interpolated images has also been proposed. The proposed approach avoids the expensive global optimization typical of optical flow methods and is easy to implement and parallelize. This allows frame interpolation at a fraction of the computational cost of traditional optical flow-based solutions.\newline

Recently, phase-based methods have shown promising results in applications such as motion and view extrapolation \cite{didyk2013joint}\cite{wadhwa2013phase}. These methods are based on the assumption that small movements of phase information can be encoded  in a very limited way, limiting their practical applicability.\newline

To solve this basic problem, \textbf{Meyer et al} \cite{meyer2015phase} proposed a method of propagating phase information across aligned multiscale pyramid levels using a new bounded shift correction strategy. The proposed algorithm estimates and adjusts phase shift information using a coarse-to-fine approach, assuming movement as in low-frequency content, in high-frequency content in a similar fashion. In this method, an adaptive phase shift upper bound is proposed that effectively avoids artifacts in large motion and an extension of phase-based imaging that results in smoother transitions between interpolated images.\newline

But the problem becomes more challenging when it comes to a light field video for the following reasons:\newline

 1. The frame rate of the light field video can be extremely low.\newline

 2. 4D light fields rather than 2D images need to be considered.\newline
 
 To super-resolve the temporal dimension of light field, \textbf{Wang et al} \cite{wang2017light} developed a hybrid imaging system consisting of  Lytro Illum \cite{Lytro} and a DSLR camera. This hybrid imaging system captures temporal resolution by incorporating other popular digital video cameras. By using a \textbf{3fps} light-field collection and \textbf{30fps} popular  2D video, the proposed system can produce a complete light field video at \textbf{30fps}. They followed a learning-based approach that could be decomposed into steps: \textbf{Spatio-temporal flow estimation} and \textbf{Appearance estimation}. Flow estimation passes angle information from the light-field collection to the 2D video so that the input image can be warped to the target view. The appearance estimation is then used to combine the warped images and output the final pixel. The whole process is trained end-to-end by using convolutional neural network. Therefore, in summary,a CNN architecture is proposed to combine a light field and 2D videos. In particular, a disparity CNN and an optical flow CNN are trained without utilizing ground truth, and cascade them to combine the angular and the temporal information.
 
 \subsection{\textbf{Spectral Reconstruction}}Rather than directly acquiring hyperspectral light field at the expense of resolution in other dimensions, a practical approach is to divide the task into 4D light field imaging and hyperspectral imaging  and then  5D light field reconstruction.\newline
 
 A novel approach is proposed by \textbf{Wu et al} \cite{wu2016snapshot} \textit{i.e.} \textbf{Snapshot for Hyperspectral Volumetric Microscopy} (SHVM). It provides a new tool for recording momentary 4D information of dynamic samples that have never appeared before. The first is a combination of an absorptive microfilter array and microlens array for 4D  spatially-spectrally coupled scanning, but the data throughput of a single sensor is insufficient to provide 5D data. For this reason, in this method,a camera array-based light field microscope is proposed to build a 4D sensor with a large instantaneous data throughput capability (approximately 1778MB / sec for RGB sensors). Data multiplexing is introduced by placing various broadband color filters in front of the sensor array, enabling 4D scanning with high light efficiency. Next, each captured sensor image is a spatially-spectrally coupled point spread function (PSF) convolution that corresponds to the hyperspectral volume data. A new 4D deconvolution algorithm has also been proposed to reconstruct high-resolution hyperspectral volume data from spatially-spectrally coupled scans without forcing strong assumptions about microscope samples.\newline
 
 \textbf{Cui et al} \cite{Cui:20} presented a  hyperspectral light field imaging system for single camera snapshots. By integerating an unfocused light field camera with a snapshot hyperspectral imager, the image mapping spectrometre, they captured a five dimensional(5D) data cube in a single camera exposure. The corresponding volumetric image (x,y,z) at every wavelength is then computed through a scale depth transform. Therefore, this method offers a snapshot hyperspectral light field imaging system that integrates light field imaging with image mapping spectrometry. Therefore,the resultant system can capture spatial, angular, and spectral records of the light field in a single snapshot and with a single camera, measuring multidimensional information i.e. a data cube of size 66 x 66 x 5 x 5 x 40.(x,y,u,v,$\lambda$)\newline
 
 The \textbf{Compressive Spectral Imaging (CSI)} system is a new era of spectral imagers because it records all wavelengths of the scene in one snapshot. 3D range imaging, on the other hand, records a 2D image of the scene in which, each pixel represents a measurement of the distance from the camera sensor to the target surface. For example, a time-of-flight (ToF) range camera indirectly measures the round-trip propagation delay of the light signal by strategically triggering the exposure time of each pixel to capture a portion of the projected pulse. In some cases,  information from both imaging modalities.\newline
 
 \textbf{Rueda et al} \cite{rueda2017spectral+} introduced a new multispectral + ToF imaging camera that simultaneously measures and reconstructs multispectral and depth (MS + D) images. The ToF sensor's dual-mode reading of ambient and modulated light is used to jointly extract the characteristics of the target scene along with depth, spatial, and spectral dimensions. The proposed camera achieves reconstruction with a spatial resolution of 128 x 128 pixels, seven spectral channels, and centimeter-level depth estimation.
 
 \subsection{\textbf{Miscellaneous}} In 2019, \textbf{Cheng et al} \cite{cheng2019light}\cite{Lightfieldtech} categorized  existing super-resolution techniques into Projection-Based, Optimization-Based and Learning-Based.\newline
 
 \textbf{Projection Based}: \textbf{Liang et al} \cite{liang2015light} developed a light transport framework to understand the basic resolution limitations of light field cameras. In this framework a pre-filtering model for a lenslet-based light field camera is derived. The main novelty of the proposed model is  the consideration of the complete space angle sensitivity profile of the optical sensor. In particular, real pixels have non-uniform angular sensitivity and respond to light along the optical axis. They have shown that a complete sensor profile plays an important role in defining the performance of a light field camera \cite{liang2015light}. The proposed method allows us to model all existing lenslet-based light field cameras and compare them in a consistent manner in a simulation, regardless of the actual differences between specific prototypes. They went further and analyzed the performance of two rendering methods, a simple projection-based method, and the inverse light transport process \cite{liang2015light}.\newline
 
 Plenoptic cameras use a microlens array to measure the radiance and direction of all the light rays in the scene. It consists of $n$ x $n$ microlenses, each of which creates an image of $m$ x $m$. Previous approaches to  depth and all-in-focus estimation problems processed plenoptic images, generated $n$ x $n$ x $m$ focus stack, and obtained a $n$ x $n$ depth map and all-in-focus image of the scene. This is a major drawback of 3DTV's plenoptic camera approach. Since the total resolution of the camera, $n^2m^2$ is divided by $m^2$ to get the final resolution of $ n^2 $ pixels.\newline
 
 \textbf{Nava et al} \cite{nava2009simultaneous} proposed a new method for simultaneously estimating both depth maps and all-in focus images of a scene from a plenoptic camera at super-resolution. Therefore, a new super-resolution focal stack combined with a multi-view depth estimation method has been proposed. With this technique, the theoretical resolution is about $n^{2}m^{2}/4$ pixels.\newline
 
 \textbf{Optimization Based Methods}: The methods in this category already discussed before.\newline
 
 \textbf{Learning-Based Methods}: \textbf{Fan et al} \cite{fan2017two} investigated a convolutional neural network approach for light field super-resolution in which the prior distribution of the image can be embedded in the CNN and are motivated by the assumption that both external and internal correlations are essential in the Light Field Super-Resolution. Since LF images are actually natural images except for angular resolution,  external correlation helps to super-resolve a single image from a general collection of images, while internal correlation is essential to enhance a single view in LF with the details in the other view. Therefore, in this method, a two-step CNN is proposed. In this case, the two steps utilize external and internal correlations, respectively. In addition, to improve CNN's second-stage generalization capabilities for inter-view super-resolution,various patch-level views have been aligned to compensate for disparities essential for LFSR. Therefore, the second stage is called a \textbf{multi-patch fusion CNN}.The first stage reuses the CNN for the general image SR to utilize the external correlation, and the second stage uses the internal correlation via designed CNN to perform the inter-view SR. In the second stage, in order to deal with the disparities between the views, this method performed view registration and aligned them before joining them. View registration is done at the patch level. For each patch in the super-resolved target view, similar patches are found in adjacent views to form an aligned view. Therefore, the CNN of the inter-view SR is called the \textbf{Multipatch Fusion CNN} (MPCNN).\newline 
 
 \textbf{Gul et al} \cite{gul2018spatial} presented a learning-based approach to light field enhancement. Both spatial and angular resolution of the captured light field is enhanced using convolutional neural networks. This method has two subnetworks. One is trained to increase the angular resolution, that is, to synthesize novel viewpoints (sub-aperture images). The other is trained to increase the spatial resolution of each sub-aperture image. The proposed method significantly improves image quality visually and quantitatively and improves the accuracy of depth estimation in terms of peak signal-to-noise ratio and structural similarity index.\newline
 
 \textbf{Yuan et al} \cite{yuan2018light} proposed an SR method that combines the Deep Convolutional Neural Networks (CNN) framework to obtain high-quality, geometrically consistent light-field images. The spatial resolution of sub-aperture images is individually enhanced by single-image super-resolution deep CNN. Next, an epipolar plane image enhancement deep CNN is proposed to restore the geometrical consistency of these images.\newline
 
 \textbf{Ivan et al} \cite{ivan2020joint} showed that both super-resolution problems can be solved together from a single image by proposing a single end-to-end deep neural network that does not require a physical-based approach. Two novel loss functions have been proposed based on the knowledge of the known light-field domain to allow networks to consider the relationships between sub-aperture images. Therefore, a new joint deep neural network for light field spatial and angular  SR has been developed that uses appearance flow to synthesize new views. Spatio-Angular Consistent loss function based on known light-field domain knowledge has also been proposed to help the network learn robustly and efficiently.\newline
 
 \textbf{Jin et al} \cite{jin2020light} proposed a new learning-based LF spatial SR framework. The framework first super-resolves each view of the LF image individually by examining the complementary information between the views in which the combinatorial geometry is embedded. A regularization network trained via the structure-aware loss function is added to accurately preserve the parallax structure along with the reconstructed view, and the correct parallax relationship is applied throughout the intermediate estimation. Therefore,a learning-based method of  LF spatial SR has been presented, addressing two important issues \textit{i.e.} how to make full use of complementary information between views and how to preserve the LF parallax structure in reconstruction. By modeling them with two subnetworks \textit{i.e.} \textbf{All-to-One SR} via \textbf{combinatorial geometry embedding} and \textbf{structural consistency regularization}.\newline 

\section{Conclusion}
In this article, we provided a critical review of Light field Super-resolution methods, starting from light field acquisition methods, via challenges associated with different acquisition methods to light field super-resolution. The light field acquisition section focuses on existing devices or methods for light field acquisition. Present light-field acquisition methods can be broadly classified into three fundamental categories i.e. multi-sensor capture approach, time-sequential capture approach, and multiplexed imaging.\newline     
Lytro and Raytrix \cite{Lytro} have commercialized light field capture devices. Other companies such as \textbf{FOVi3D} \cite{fovi3d} and \textbf{Japan Display} \cite{JDI} also manufacture light field projection solutions. Unlike Lytro cameras, which place the microlens array in front of the image sensor, \textbf{Wooptix} \cite{woop} uses a liquid lens \cite{kuipers2004variable} in the optical chain in front of the sensor to quickly switch focal planes, thereby achieving full resolution in real-time with reduced resolution trade-off.  In addition, \textbf{Google} has published a number of patents related to light-field capturing \cite{ng2018depth}. They also designed the camera to capture light field images with non-uniform and incomplete angular sampling. As a result, not only the spatial resolution but also the quality of the depth data has improved \cite{pitts2018capturing}. On light field displays, \textbf{Avegant} \cite{avegant}, \textbf{Leia} \cite{leia}, \textbf{Light Field Lab} \cite{Lightfieldlab}, \textbf{Dimenco} \cite{dimenco}, and \textbf{Creal} \cite{creal}  enable realistic digital photography on the screen. Similarly, the \textbf{Looking Glass Factory} \cite{look}  is a light field image that offers 45 different viewing angles as long as the viewer is in a 58-degree field of view, is much more cost-effective than previous products and hence more attainable. In 2020, Sony released two light-field-related technologies i.e. \textbf{3D Spatial Reality Display} \cite{spatial} and \textbf{AtomView} \cite{atomview}. The SR Display recreates  3D spatial images  pretends like as if they were real, and can be seen with the naked eye without special glasses or headsets. It also achieves a relatively high resolution. Atom View technology revolutionizes the use of high-quality photorealistic volume data for use in virtual production and entertainment experiences. The combination of unparalleled real-time visual quality and built-in streamlined workflow tools allows developers to focus on telling a unique and impactful story.\newline

Researchers have paid much attention in light field super-resolution because of the natural resolution trade-off associated with light-field acquisition systems\cite{wu2017}. For individual users, the capturing process is highly random and requires stabilization, with a high-resolution display like a regular camera. \cite{Lightfieldtech}.Therefore, from the above discussion, the following conclusion have been drawn i.e.\newline

1. Light-field super-resolution can greatly benefit from hybrid imaging systems. These have several advantages, including: Much smaller data size and compressibility\cite{wu2017}.\newline

2. Learning-Based approaches (such as CNN and sparse coding) can improve performance, especially when taking advantage of the properties of the light-field structure such as an EPI structure\cite{wu2017}.\newline

3. Super-resolution of non-Lambertian regions can be a major challenge, especially for a light-field microscopy where depth information is difficult to obtain\cite{wu2017}.

\bibliographystyle{plain}
\bibliography{root}

\end{document}